# Inhibition of water vapor condensation by dipropylene glycol droplets on hydrophobic surfaces via vapor sink strategy

Jun Hu[1,2], Zhan-Long Wang[1]*

**AFFILIATIONS**

[1]Shenzhen Institute of Advanced Technology, Chinese Academy of Sciences, Shenzhen, Guangdong 518000, China.

[2]China University of Political Science and Law, Beijing, 100091, China.

*Author to whom correspondence should be addressed: zl.wang1@siat.ac.cn

Abstract

Condensation, frosting and icing are natural phenomena, and have been enduring challenges for human society and modern engineering. These phenomena pose a range of issues, from hazardous icy road surfaces, damage to electronic devices due to condensation, to frost accumulation on power lines and aircraft. Dipropylene glycol (DG) is a non-toxic, non-corrosive, and biologically safe substance with excellent hygroscopicity, capable of inhibiting condensation and ice formation through the "vapor sink" strategy. However, there has been limited research on DG. Herein, we provides a detailed study of the performance of DG droplets in suppressing condensation with experiments and theoretical analysis. Results found that a dry zone ring would form around the DG droplet on a cooling solid surface. The ratio of the dry-zone radius to the DG droplet radius increases with the solid surface temperature and scales as $1/(T_{\text{dew}}-T_{\text{c}})$. Results indicated that the cooling time of substrate surface does not affect the ratio in a short period. When the temperature is higher, the ratio decreases slightly with DG droplet volume; while it keeps almost still, when the temperature is lower. A theoretical model is also proposed to reveal the relationship between the ratio and the substrate surface temperature. These findings hold the promise of not only enhancing our understanding of condensation suppression but also advancing the development of innovative solutions for various industries, including transportation, electronics, aviation, and power distribution.

# 1. Introduction

Condensation, and consequently frost formation and ice accumulation are ubiquitous phenomena encountered in diverse environments, giving rise to a plethora of challenges in both engineering and everyday life. The development of dew or frost not only impairs visibility and road safety but can also inflict significant damage on electronic devices, compromising their functionality and longevity. Furthermore, in critical sectors such as aviation and power distribution, frost formation on aircraft surfaces and power lines poses safety risks and has the potential to disrupt power transmission[1-5]. As a result, the development of effective anti-condensation and consequently frost and ice formation strategies holds immense importance for the electrical industry and various applications, including power harvesting and self-cleaning surfaces[6-9]. Strategies to tackle the challenges of anti-icing and anti-frosting encompass a range of approaches, including inhibiting ice nucleation, retarding frost formation, reducing ice adhesion, and minimizing frost accumulation[2,10-15]. Prior research has explored various methods, including the study of droplet self-jumping during condensation, the comprehension of spontaneous droplet motion[16-19], the design and fabrication of superhydrophobic surfaces or surfaces with specific structures[20-22], and the utilization of hygroscopic materials to inhibit vapor condensation[10,23].

Inhibiting the condensation of a hygroscopic liquid, known as a hygroscopic vapor sink, is an important method for anti-condensation, anti-frosting and anti-icing[10,24-26]. In recent years, scientists have designed surfaces based on this concept to suppress the formation of condensation, frosting and icing. For example, since ice has a lower vapor pressure than liquid water, Ahmadi and colleagues designed a passive anti-frosting surface using microscopic ice stripes to suppress the in-plane growth of frost[27]. Sun and co-workers demonstrated that an array of hygroscopic droplets on solid surfaces could create large areas that inhibit condensation and suppress frost and ice formation. They designed a two-layer structure consisting of glycol and a porous thin film[23,28]. Despite the promising applications of hygroscopic liquids in anti-icing and anti-frosting, and the recent advances, there are still challenges and significant potential for development in related research on hygroscopic materials and their anti-condensation properties. Additionally, based on this research, there is potential to develop effective dehumidifying materials and materials that inhibit the condensation of water molecules.

Hygroscopic materials play a crucial role in preventing the condensation of water vapor molecules, creating a dry annular region referred to as a dry zone. This dry zone is effectively depicted in Fig. 1, which includes both vertical (Fig. 1a) and side (Fig. 1b) views of the dry zone formed around the hygroscopic droplet. In addition to these representations, Figure 1c and 1d provide illustrations of condensation (Fig. 1c) and icing (Fig. 1d) scenarios, highlighting the dry zone formation resulting from the hygroscopic nature of the droplet. The significance of these findings extends to potential applications in anti-icing and anti-frosting functionalities in emerging smart films and various materials. Understanding the consistent efficacy of different hygroscopic materials in suppressing condensation is of paramount importance as it serves as a guiding principle for the development of innovative anti-condensation materials.

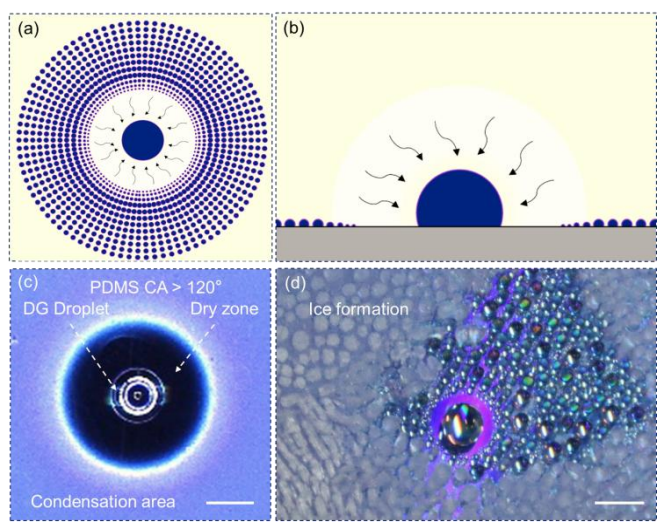

**Fig. 1.** The diagram of anti-condensation of DG droplet due to the hygroscopicity. (a)-(b) The vertical (a) and side (b) views of the diagram of anti-condensation droplet and the dry zone formation; (c) The experimental image of anti-condensation properties of hygroscopic droplet on PTFS; (c) The inhibition of icing due to the anti-condensation properties of hygroscopic after long time cooling.

However, until now, the majority of research on materials inhibiting condensation has been confined to salt aqueous solutions, hydrogels, fibers, and similar substances. Examples of such materials encompass cellulose fibers[29], salts, saturated salt solutions (e.g., NaCl crystals and lithium bromide solution)[30,31], alcohols (such as and glycerol)[32], silica gels[33], hydrogels[11,34,35], and ice[26]. Guadarrama-Cetina et al. conducted observations of the formation of a dry zone around a NaCl solution droplet, studying the water vapor pressure above the droplet. Their findings revealed a decrease in the water vapor pressure above droplets with different concentrations, compared to the saturated water vapor pressure[30,31]. Additionally, owing to osmotic pressure, hydrogels exhibited hygroscopic abilities. Delavoipière et al. measured the transfer of vapor from the atmosphere to hydrogel thin films and explored the impact of film hygroscopicity on the dynamics of film swelling[34]. Sun et al. demonstrated that glycol could effectively inhibit vapor condensation. They designed a bilayer-coated structure with a porous exterior covering a hygroscopic liquid-infused layer to control vapor concentration[23,28]. Furthermore, ice has been utilized to inhibit vapor condensation due to its lower saturated vapor pressure compared to water[27].

As a hygroscopic liquid, dipropylene glycol (DG) exhibits exceptional proficiency in suppressing water vapor condensation[36,37]. DG has garnered significant attention in this regard and is known for its low toxicity, high stability, and excellent solubility. With a chemical formula of $C_3H_8O_2$, it finds extensive applications across various industries. In the industrial and manufacturing sectors, DG serves as a versatile solvent, lubricant, and thickening agent. In food processing, it acts as a sweetener, preservative, and stabilizer. Furthermore, in the pharmaceutical industry, DG functions as both a solvent and lubricant for medications. Beyond this, DG is a key component in the production of coatings, inks, dyes, and fragrances, among other products. Its versatility and widespread availability position DG as a highly promising compound with the potential to play a vital role in diverse domains. The unique hygroscopic properties of DG make it an intriguing candidate for addressing challenges related to condensation. Through active moisture absorption from the surrounding environment, it can effectively mitigate the onset of condensation, frost, and ice formation. However, despite its immense potential, research on the practical applications of DG in this context remains relatively underexplored.

In this paper, we aim to bridge this knowledge gap by conducting a detailed investigation into the efficacy of DG droplets in the context of condensation inhibition. Through systematic experimentation and analysis, we seek to uncover the underlying mechanisms and potential applications of DG as a potent anti-condensation agent. The findings from this study hold the promise of not only enhancing our understanding of condensation suppression but also advancing the development of innovative solutions for various industries, including transportation, electronics, aviation, and power distribution.

## 2. Materials and methods

To illustrate this phenomenon, we designed specialized equipment capable of capturing droplet condensation dynamics. This equipment comprises a capture platform for monitoring the evolution of condensation droplets and a cold source responsible for generating these condensed droplets, as depicted in Fig. 2a. Our experimental investigations were carried out on a hydrophobic surface constructed from a 100 μm-thick polydimethylsiloxane (PDMS) thin film. This PDMS thin film was created by mixing Dow Corning materials at a mass ratio of 10:1 for the matrix and curing agent. The film had dimensions of 2 × 2 cm$^2$. We carefully placed a droplet of DG (sourced from Shanghai Aladdin Biochemical Technology Co., Ltd.) at the central point of the PDMS thin film surface (PTFS). The contact angle of water droplets on PTFS measured approximately 120º, while for DG droplets on the PDMS surface, the contact angle was recorded as 95°. Consequently, in this particular case, vapor condensation occurred in a dropwise manner. The PTFS is smooth enough, avoiding the effect of sharp surface roughness on the condensed droplet distribution[38]. To ensure an even temperature distribution across the PTFS, we affixed the PDMS film onto a custom-cut heat conduction strip. This strip was positioned atop a semiconductor cooling chip (SCC, Model12715). We further enhanced our experimental setup by installing an infrared thermal imager to continuously monitor the temperature of the PTFS. The SCC was controlled using a DC power source, with a maximum applied voltage of 12 V. Temperature regulation was achieved by adjusting the voltage supplied by the DC power source. This setup allowed us to maintain a constant temperature on the PDMS thin film for an extended period, ensuring the conditions required for our experiments. To dissipate the heat generated by the SCC, we connected the heating side of the SCC to a cooling circulation system (CCS). The experimental setup was placed in a temperature and humidity chamber to keep stable ambient temperature and relative humidity (RH), as shown in Fig. 2a.

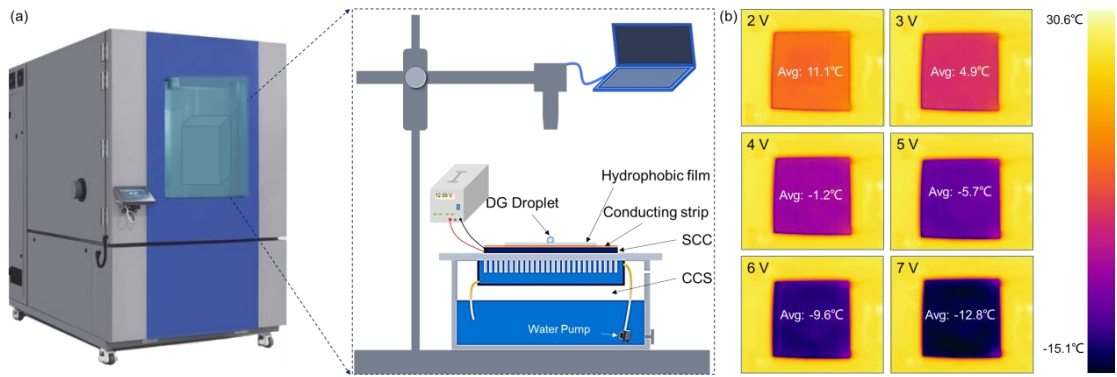

**Fig. 2.** Experimental setup and temperature control: (a) The experimental setup included a CCD, hydrophobic PDMS thin film, light disk, SCC, CCS, and DC power source. A transparent chamber was used for the devices to stabilize the experiments and for ease of manipulation; (b) Temperature distribution image on SCC. As the voltage decreases, the average temperature of conducting strip surface on SCC reduces. When the power source was turned on, the temperature decreased rapidly and then stabilized approximately several tens of seconds later.

The temperature of the system was controlled by varying the voltage applied to the SCC. An investigation was conducted to determine how changes in the input voltage affected the surface temperature of PTFS on the SCC, as shown in Fig. 2b. The voltage parameters ranged from 2 V to 7 V, with increments of 1 V used to ensure robust results. Data were collected using an infrared thermal imager and analyzed based on the average temperature detected on the PTFS. Initially, the SCC was subjected to an input signal of 2 V, resulting in a surface temperature measurement of 11.1°C. Subsequently, the applied voltage was increased by 1 V at each stage, and temperature measurements were taken. At 3 V, the average temperature had decreased to 4.9°C. This decreasing trend continued as the voltage increased further: -1.2°C at 4 V, -5.7°C at 5 V, -9.6°C at 6 V, and finally stabilizing at -12.8°C when a voltage of 7 V was applied. Infrared images clearly depicted a consistent decline in the average surface temperature as the voltage increased. These observations illustrated a strong inverse relationship between the applied voltage and the resulting surface temperature of the SCC. As the voltage was incrementally raised, the average surface temperature significantly decreased. The temperature decrease was rapid immediately after turning on the power and typically took less than one minute. Following this initial drop, the temperature remained relatively stable, with temperature fluctuations of less than 0.1°C. This ensured precise temperature control of the film surface during the experiments. The temperature will exhibit a tendency to initially decrease and then increase after reaching a transition point with increasing voltage. This phenomenon occurs because, once the applied voltage exceeds a critical value, the heat generated by the SCC cannot dissipate rapidly enough, thereby inhibiting further temperature reduction.

**3. Results and discussion**

3.1 *The variation of the dry zone around DG droplet*

The anti-condensation properties can be characterized by the formation of a dry zone, in which the condensation of water vapor molecules is inhibited. This behavior of DG droplets with different volumes is illustrated in Figs. 3 and 4, with specific emphasis on two cases. In Fig. 3, we focus on a 2 μl droplet. Here, we examine the changes in the dry zone area concerning temperature and discuss the results. The substrate surface temperatures in Fig. 3a, from left to right, are 11.1 °C, 4.9 °C, -1.2 °C, -5.7 °C, -9.6 °C, and -12.8 °C. We provide experimental snapshots that display the variation of the dry zone at different temperatures while maintaining a constant relative humidity (RH). The room temperature and ambient RH were set at 25 °C and 60%, respectively. The dew point temperature can be determined either through empirical formulas or by employing a temperature and humidity meter. In our experiments, the dew temperature $T_{dew}$ is measured at 15.8 °C. Our results reveal that a dry zone forms around the DG droplet, taking on a ring-like shape. This phenomenon can be attributed to the hygroscopic nature of DG, where the molecules in the DG liquid possess a strong capacity to capture water vapor molecules. Consequently, this leads to a lower concentration of water vapor molecules around the DG droplet. Furthermore, the RH closely associated with water vapor molecules around the DG droplet is lower than what is necessary for condensation to occur. This makes condensation unlikely in the vicinity of the DG droplet. In regions farther away from the DG droplet, the RH is minimally affected by the hygroscopic effect, allowing normal condensation to occur as the temperature decreases. Given that the droplet is spherical, the dry zone area takes on the form of a ring around the DG droplet. Further discussion of this mechanism is provided in the Theoretical Analysis section in the following part of this study.

In Fig. 3a, we observe the formation of dry zone area rings around the DG droplet for all the temperatures studied. The $T_{dew}$ is 15.8 °C. In the images presented in Fig. 3a, the temperatures are consistently lower than $T_{dew}$, indicating the occurrence of condensation on the PTFS. In all cases, the dry zone area is largest at the highest temperature of 11.1 °C, as illustrated in the first image. From left to right and from top to bottom, there is a noticeable reduction in the dry zone area as the temperature decreases from 11.1 °C to -12.8 °C. This decrease can be attributed to the hygroscopicity of DG droplets, which results in a non-linear decrease in the concentration of water vapor surrounding the droplets. This phenomenon is depicted in the dotted line in Fig. 3b. As it gets closer to the DG droplets, the dew point temperature decreases. As the PTFS temperature decreases, the boundary of the condensation area approaches the DG droplet. Figure 3b presents two cases for comparison. In case 1, the temperature of PTFS is denoted as $T_1$, while in case 2, the temperature is $T_2$. Condensation occurs when the PTFS temperature is less than or equal to the dew temperature. As a result, in case 1, the condensation area edge is located at point $L_1$, while in case 2, it is at point $L_2$. Given that $T_1 > T_2$, the boundary of the condensation area in case 1 is farther away from the droplet, resulting in a larger dry zone area surrounding the droplet when compared to case 2. In Fig. 3c, the radii of dry zone and DG droplets at different temperatures are shown. From the results, the radius of the dry zone decreases as the PTFS temperature decreases. Conversely, the radius of DG droplets remains relatively constant across different temperatures. This consistency can be attributed to the fact that, over a short period, the flux of water vapor absorbed by the surface of DG droplets does not exhibit significant variation at different temperatures.

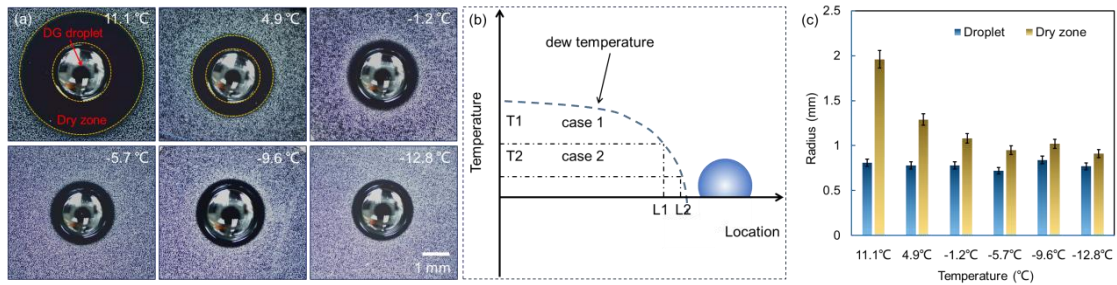

**Fig. 3.** The anti-condensation of DG droplets at different temperatures. (a) The dry zone variation at different temperature. The temperatures are 11.1 °C, 4.9 °C, -1.2 °C, -5.7 °C, -9.6 °C and -12.8 °C, respectively. Obviously, the dry zone area reduces with decreasing temperature. (b) The diagram of the effect of temperature on the dry zone area. (c) The variation of the radii of the droplet and the dry zone at different temperatures. The radius of dry zone decreases with decreasing temperature. The radius of droplet keeps almost the same for the same cooling time at different temperatures.

Figure 4 illustrates the behavior of a 4 μl DG droplet under the same PTFS temperatures as shown in Fig. 3. In Fig. 4a, the formation of the dry zone around the DG droplet follows a similar trend as seen in Fig. 3. As the temperature decreases, the dry zone area also decreases. Figure 4b provides data on the radii of the dry zone and DG droplet at various temperatures. These results confirm that a decrease in temperature leads to a reduction in the dry zone radius, while the radius of DG droplet remains relatively unchanged.

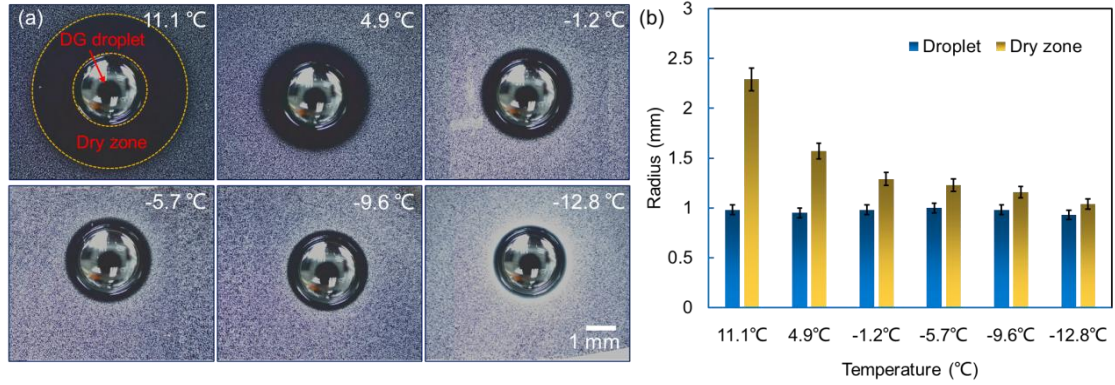

**Fig. 4.** The anti-condensation of a 4-μl DG droplet. (a) The images of the situation of condensation around DG droplet. The temperatures in different images from the left to right and up to bottom are 11.1 °C, 4.9 °C, -1.2 °C, -5.7 °C, -9.6 °C and -12.8 °C, respectively. The dry zone area obviously reduces with dropping temperature. (b) The radius variation of dry zone and DG droplet with temperature. The dry zone radius decreases as the temperature reduces.

3.2 *The effect of substrate surface temperature on ratio RA*

To investigate the influence of experimental parameters on the variation of the dry zone, we introduced a dimensionless parameter, denoted as RA that is defined as the ratio of the radius of the drying zone to the radius of the DG droplet. In Fig. 5, we present the variations of RA with respect to the PTFS temperature. We explore four different cases, each involving droplets of varying volumes and cooling times: 1 μl (Fig. 5a), 2 μl (Fig. 5b), 3 μl (Fig. 5c), and 4 μl (Fig. 5d). Our dataset encompasses various cooling times, including 2 minutes, 3 minutes, 4 minutes, 5 minutes, 6 minutes and 7 minutes, since the power source was turn on. The results reveal a nonlinear increase in RA with rising temperature. As the temperature escalates, the rate of change of RA exhibits a gradual amplification. When the temperature remains below 0 °C, both the slope and the amplitude of the RA alteration are relatively small. Furthermore, as the temperature continues to decrease, the RA experiences only minor fluctuations. According to previous reports[30], the concentration of water vapor follows a hyperbolic function. Specifically, the dry zone radius scales with $1/(T_{dew} - T_c)$, where $T_{dew}$ denotes the dew temperature, and $T_c$ represents the substrate surface temperature. A comprehensive discussion of this phenomenon can be found in the Theoretical Analysis section. Figure 6 further illustrates the variations of RA concerning temperature, showcasing data at various cooling times: 2 minutes (Fig. 6a), 3 minutes (Fig. 6b), 4 minutes (Fig. 6c), 5 minutes (Fig. 6d), 6 minutes (Fig. 6e), and 7 minutes (Fig. 6f). We maintain four different droplet volumes ranging from 1 μl to 4 μl. Remarkably, the trends in RA variation remain consistent across different condensation times, as shown in Fig. 5.

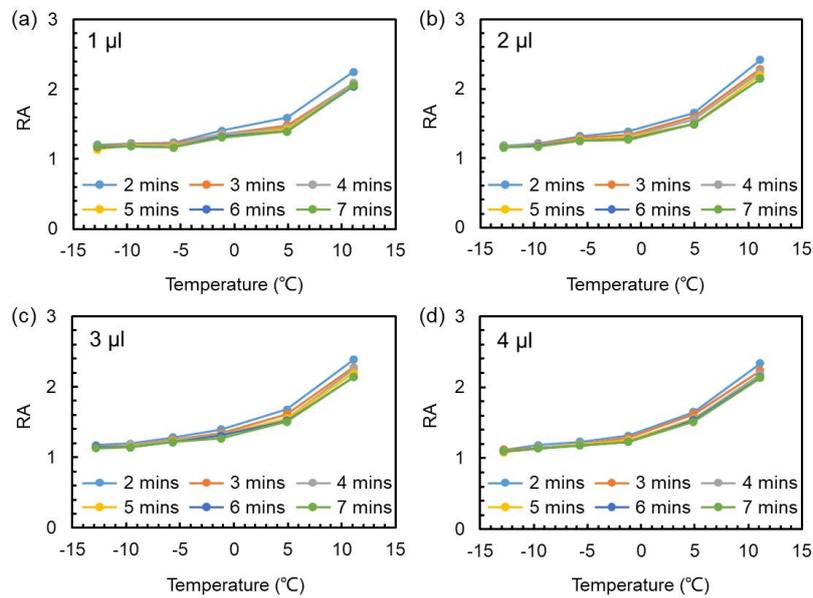

**Fig. 5.** The variation of RA with temperature for fixed cooling time. (a)-(d) show the cases of 1 μl (a), 2 μl (b), 3 μl (c) and 4 μl (d), respectively. Results indicated that RA increases with temperature nonlinearly. As the temperature increases, the slope of RA also increases.

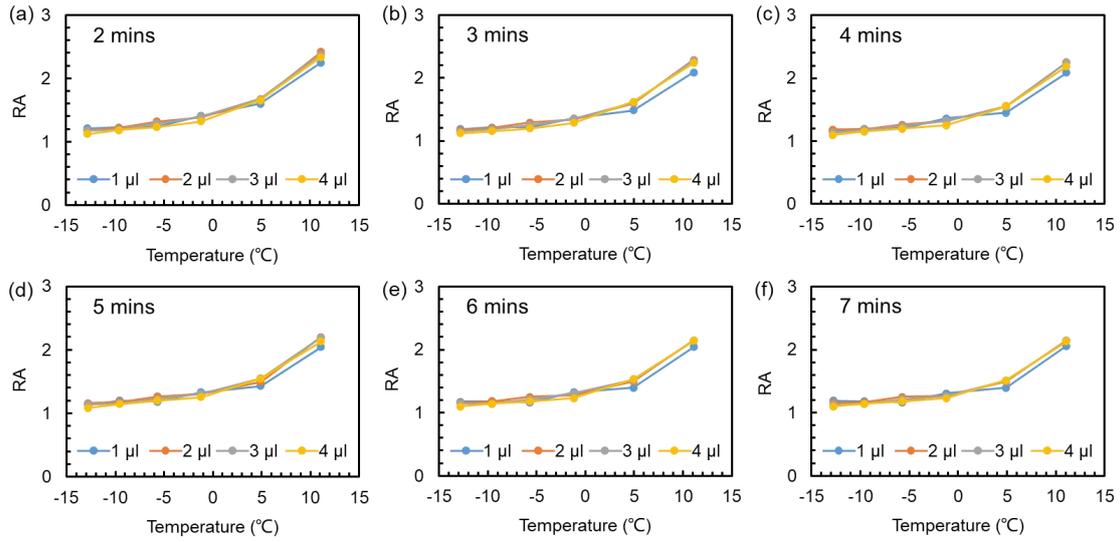

**Fig. 6**. The variation of RA with temperature for fixed cooling time. (a)-(f) show the cases of 2 mins (a), 3 mins (b), 4 mins (c), 5 mins (d), 6 mins (e) and 7 mins (f), respectively. RA increases with the temperature in a hyperbolic function.

3.3 *The effect of cooling time on ratio RA*

The influence of cooling time on the ratio RA is examined in Figs. 7-8. In Fig. 7, we present experimental snapshots of two cases: 1 μl (Fig. 7a) and 3 μl (Fig. 7b) droplets, both at a temperature of -1.2 °C. In Fig. 7a, the images display the cooling time ranging from 2 to 7 minutes. The results reveals that the dry zone surrounding the DG droplet remains relatively consistent during a short period. This suggests that the dry zone remains stable under short cooling durations. In Fig. 7b, which depicts the case with a 3 μl droplet, we observe a similar pattern of dry zone variation as in the case of the 1 μl droplet. This similarity indicates that, under a specific temperature condition, the moisture absorption capacity of DG droplets remains relatively stable over short periods. Figures 7c and 7d present statistical data on the radii of the dry zone and DG droplet for both the 1 μl (Fig. 7c) and 3 μl (Fig. 7d) droplets. These statistics demonstrate that there are no significant changes in the radii of the droplets and dry areas between the 2 and 7-minute intervals. Therefore, the hygroscopicity of DG droplets does not notably decrease after a short period of condensation and water vapor absorption. Moreover, the diameter of the droplet also remains relatively constant during this short period, indicating that the absorption of water vapor does not have a substantial impact on the change of droplet volume.

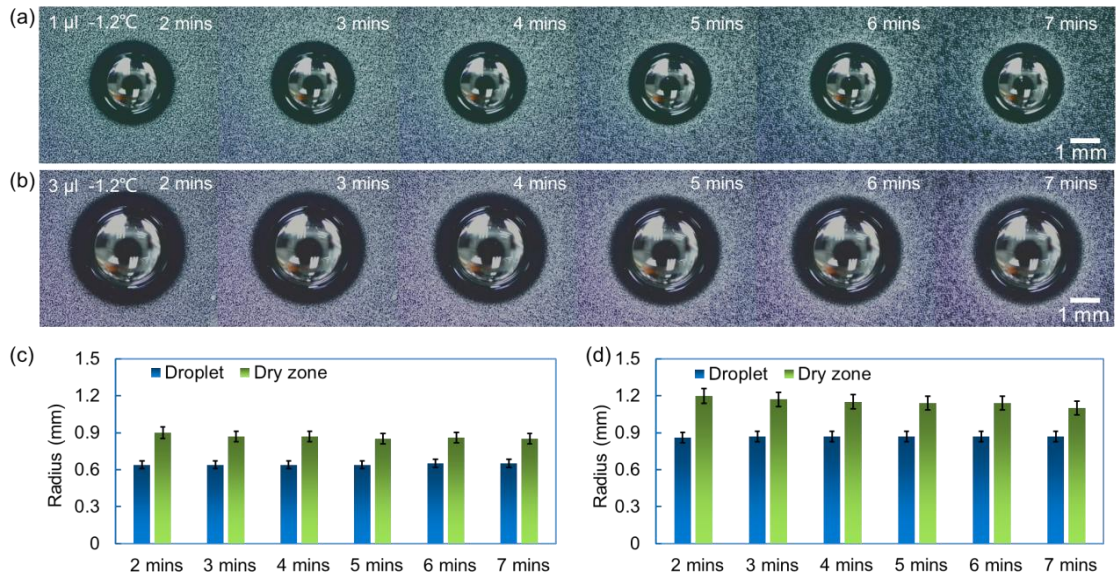

**Fig. 7.** The effect of cooling time on RA. (a) The case of 1 μl droplet at different cooling times. The PTFS temperature is -1.2 °C. The dry zone maintains almost still with the cooling time in a short period. (b) The case of 3 μl droplet at different cooling times. (c)-(d) The statistics of radii of dry zone and DG droplet at different cooling times for 1 μl droplet (c) and 3 μl droplet (d).

In Fig. 8, we observe the relationship between the ratio RA and cooling time for four different cases: 1 μl (a), 2 μl (b), 3 μl (c), and 4 μl (d). There are distinct trends for different temperature conditions. For higher temperatures, represented by cases (a) and (b), we see that the value of RA gradually decreases, albeit slightly, as condensation time increases. This suggests that as the cooling time progresses, the ratio RA, which is likely related to the extent of condensation or some other physical property, experiences a gradual reduction. In contrast, for lower temperatures, depicted in cases (c) and (d), RA remains nearly unchanged as condensation time increases. This implies that the cooling time, within the range considered, has minimal impact on the value of RA under these colder temperature conditions. These observations highlight the dependence of the RA ratio on both the initial droplet volume and the temperature conditions, suggesting a complex interplay of factors in the condensation process that varies with different parameters.

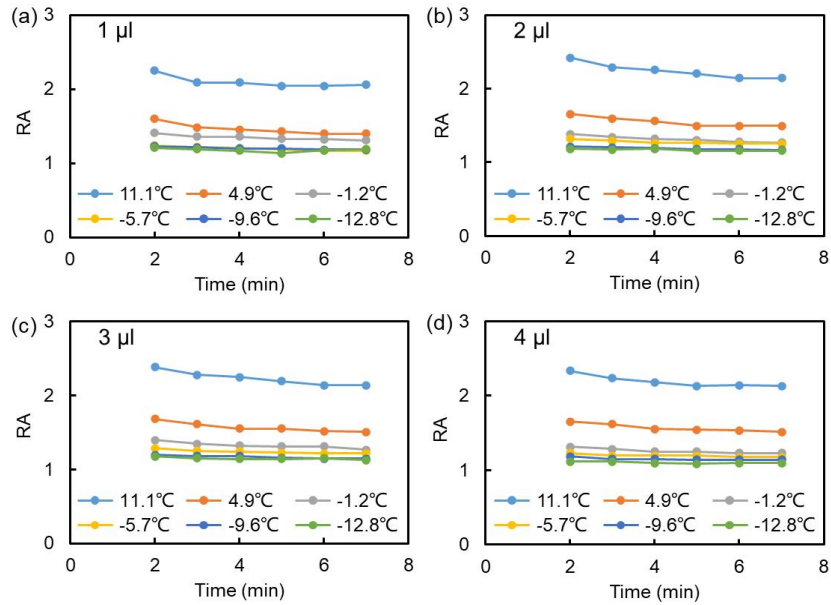

**Fig. 8.** The variation of ratio RA with cooling time. Four cases of 1 μl (a), 2 μl (b), 3 μl (c) and 4 μl (d) are shown. For higher temperatures, RA gradually decreases, albeit slightly, as cooling time increases. For lower temperatures, RA remains nearly unchanged as the cooling time increases.

3.4 *The effect of the volume of DG droplet on ratio RA*

Figures 9-11 show the variation of RA with different volumes of DG droplets. In Fig. 9, the experimental snapshots of the cases of 1 μl, 2 μl, 3 μl and 4 μl at temperatures of 11.1 °C (Fig. 9a), -1.2 °C (Fig. 9b) and -9.6 °C (Fig. 9c) are shown. From the comparison in the figure, it can be observed that the area of the dry zone increases with an increase in liquid volume. This increasing trend is more pronounced at higher temperatures, whereas it is less distinct at lower temperatures. Specifically, at a PTFS temperature of 11.1 °C, the area of the dry zone significantly increases as the droplet volume increases from 1 μl to 4 μl. However, at temperatures of -1.2 °C and -9.6 °C, the dry zone area around the 4 μl droplet is greater than that around the 1 μl droplet, yet the width of the formed dry zone ring does not exhibit a significant change compared to the case at 11.1 °C.

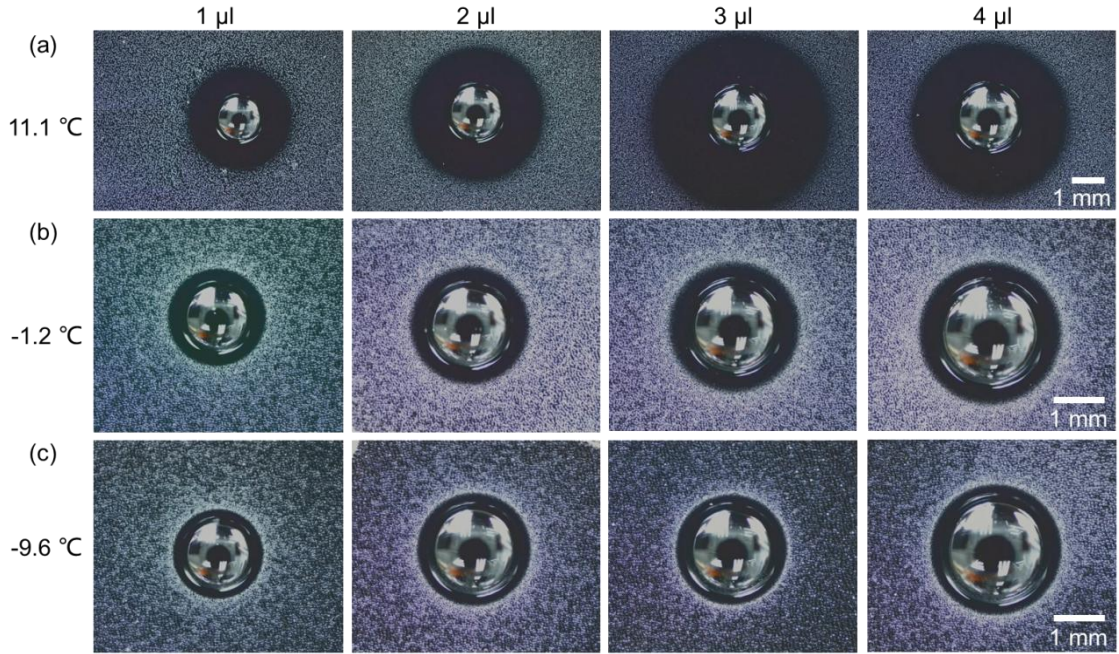

**Fig. 9.** The snapshots of the anti-condensation of DG droplet with different volumes. (a) The cases of different volumes at 11.1 °C; (b) The cases of different volumes at -1.2 °C; (c) The cases of different volumes at -9.6 °C.

Figures 10 and 11 depict the variation of the ratio RA of the dry zone area to the droplet radius with respect to the droplet volume under different temperature and cooling time conditions. In Fig. 10, subfigures 10a to 10f represent the changes of RA with liquid volume at different temperatures for 2 (Fig. 10a), 3 (Fig. 10b), 4 (Fig. 10c), 5 (Fig. 10d), 6 (Fig. 10e), and 7 minutes (Fig. 10f) of condensation time, respectively. For the same condensation time, the variation trend of RA is relatively gentle at different temperatures. Regarding different condensation times, from 2 minutes to 7 minutes, a similar variation pattern is observed. Figure 11 displays the relationship between RA and volume under different temperature conditions, namely 11.1 °C (Fig. 11a), 4.9 °C (Fig. 11b), -1.2 °C (Fig. 11c), -5.7 °C (Fig. 11d), -9.6 °C (Fig. 11e), and -12.8 °C (Fig. 11f), for various condensation times. It can be observed that at higher temperatures, RA remains relatively unchanged with volume (as seen in Fig. 10a and 10b). Conversely, at lower temperatures, RA shows a slight decreasing trend with volume. This phenomenon becomes more pronounced as the temperature decreases, particularly in subfigures 10e and 10f. This suggests that under constant experimental conditions such as temperature and humidity, for DG, the ratio of the dry zone area to the droplet radius remains unaffected by changes in droplet volume at higher temperatures. As the volume of the DG droplet increases, the area of the dry zone increases proportionally. However, at lower temperatures, the ratio of the dry zone area to the droplet radius exhibits a slow decreasing trend with an increase in volume, indicating that the dry zone area does not increase proportionally as the droplet volume increases.

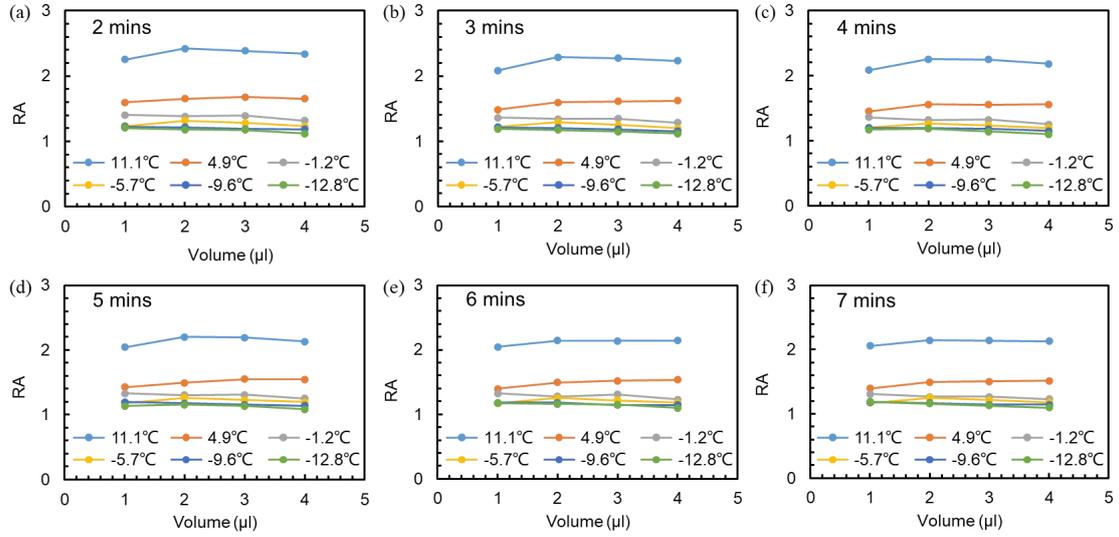

**Fig. 10.** The variation in ratio RA with DG droplet volume at different cooling times. (a)-(f) show the variation of RA at 2 mins (a), 3 mins (b), 4 mins (c), 5 mins (d), 6 mins (e) and 7 mins (f). RA maintains almost still with droplet volume at different cooling times.

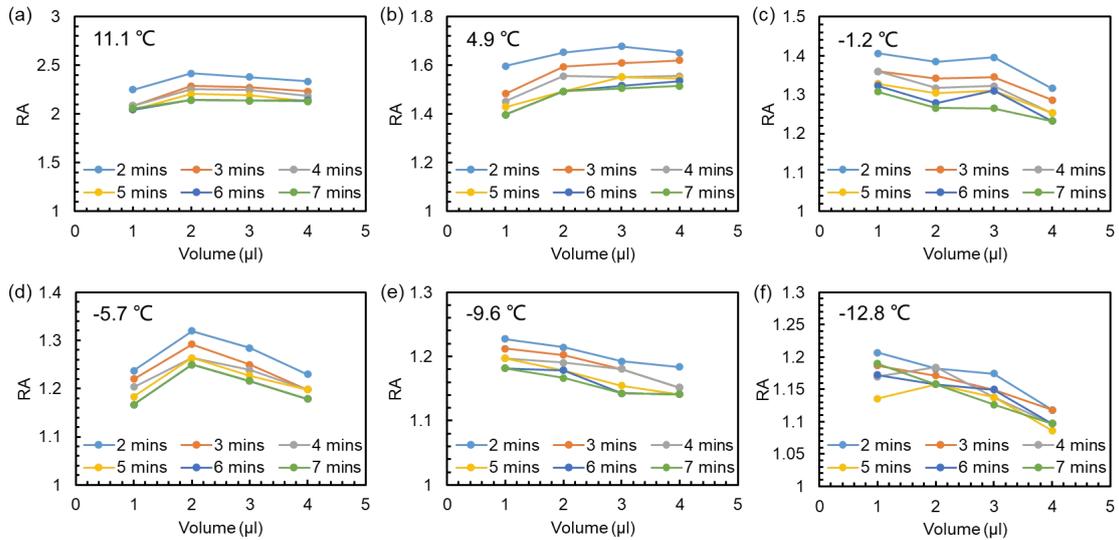

**Fig. 11.** The variation in ratio RA with DG droplet volume at different PTFS temperatures. (a)-(f) show the variation of RA at 11.1 °C (a), 4.9 °C (b), -1.2 °C (c), -5.7 °C (d), -9.6 °C (e) and -12.8 °C (f). At higher PTFS temperature, RA maintains almost still, while RA decreases slightly at lower PTFS temperatures.

## 4. Theoretical analysis

A theoretical model was developed to elucidate the mechanism governing the evolution of the dry zone. In our model, we considered the scenario where an immobile droplet with a radius denoted as 'R' was positioned on a horizontal surface. Water vapor molecules (WVMs) were diffused around the droplets. The WVM concentration $n(r, t)$ followed the dynamic equation:

$$\frac{\partial n}{\partial t} = D_0 \Delta n, \tag{1}$$

where, $D_0$ and $t$ denote the diffusion coefficient and time, respectively. $r$ is the distance to the center of the droplet. The hydroscopic droplet radius evolution follows the growth equation:

$$\frac{dV}{dt} = \int J dS. \tag{2}$$

where, $J$ is the flux of the WVMs per unit surface area, and $V$ is the droplet volume. The WVM flux can be expressed as

$$J = D_0 \frac{\partial n}{\partial r}\bigg|_R. \tag{3}$$

The evolution of the droplet and dynamic dry zone is a Stefan problem with a moving boundary at $r = R(t)$. We first discuss the influence of temperature on the variation in the dry zone. We assume sufficiently slow growth, as shown in Eq. (1), such that the time dependence of $n$ can be neglected (quasistatic approximation). We obtain the Laplace equation:

$$\Delta n = 0. \tag{4}$$

For the 3D space, Eq. (4) has a hyperbolic solution[30].

$$n = n_\infty - (n_\infty - n_R)\frac{R}{r}, \tag{5}$$

The boundary conditions are $n(R) = n_R$ and $n(\infty) = n_\infty$, where $n_R$ is the WVM concentration at the surface of the hygroscopic droplet.

For the dry-zone evolution, we considered the cooling substrate surface, the surface where condensation occurred. In this case, the distance $r$ is the length from the center of the hygroscopic droplet to the edge of the dry zone. For the variation in the dry-zone size, the distance $r$ was determined through comparison of the substrate surface temperature and dew temperature $T_{dew}$. The critical condition determining the edge of the dry zone was that the cooling surface temperature $T_c$ was equal to the dew temperature $T_{dew}$ at the edge of the dry zone, namely:

$$T_c = T_{dew}. \tag{6}$$

In general, in certain atmospheric environments, $T_{dew}$ can be calculated using the following equation, summarized by Hyland and Wexter[39]:

$$T_{dew} = \varphi(A + B \cdot T_c) + C \cdot T_c - 19.2, \tag{7}$$

where, A, B, and C are constant parameters with values of 0.1980, 0.0017, and 0.8400, respectively[39]. The WVM concentration $n$ can be characterized by the moisture content $d$. The moisture content is related to the RH $\varphi$, water vapor pressure $P_q$, saturated water vapor pressure $P_s$ ($P_q = \varphi P_s$), and local air pressure $P$ as follows.

$$n = d = 622 \frac{\varphi P_s}{P - \varphi P_s}. \tag{8}$$

Integrating, we obtain:

$$\frac{\varphi P_s}{P - \varphi P_s} = \frac{P_s}{P - P_s} - \left( \frac{P_s}{P - P_s} - \frac{\varphi_G P_s}{P - \varphi_G P_s} \right) \cdot \frac{1}{RA}, \tag{9}$$

$$\varphi \left( A + B \cdot T_c \right) + C \cdot T_c - 19.2 = T_c, \tag{10}$$

where, $\varphi_G$ represents the RH at the top of the hygroscopic droplets, and $P$ is the atmospheric pressure. $P_s$ was obtained from an atmospheric handbook. By solving Eqs. (9) and (10), we can establish a clear relationship between the ratio RA and the substrate surface temperature, denoted as $T_c$. In the dry zone, the RH is consistently lower than that in the condensation area, i.e., RH<$RH_{de}$, where $RH_{de}$ is the critical RH at the dew temperature. In contrast, in the condensation area, RH is greater than $RH_{de}$. The dew temperature changes with the RH linearly in a stable atmospheric environment.

Figure 12 illustrates the theoretical curves depicting the relationship between the ratio RA and the substrate surface temperature. In Fig. 12a, we display four $T_{dew}$ values (2 °C, 5 °C, 10 °C, and 15 °C). Upon a theoretical analysis, the variation in the ratio RA with the substrate surface temperature follows a hyperbolic function. When the temperature approaches the dew temperature from lower values, the ratio increases, ultimately tending towards infinity as it approaches $T_{dew}$. Conversely, as the temperature decreases, the ratio converges towards a specific value, determined by the properties of the hygroscopic liquid. In the temperature range between lower values and $T_{dew}$, the ratio RA experiences a gradual increase. However, as it approaches $T_{dew}$, the ratio exhibits a sharp incline. This pattern holds for various values of $T_{dew}$, demonstrating a consistent rule.

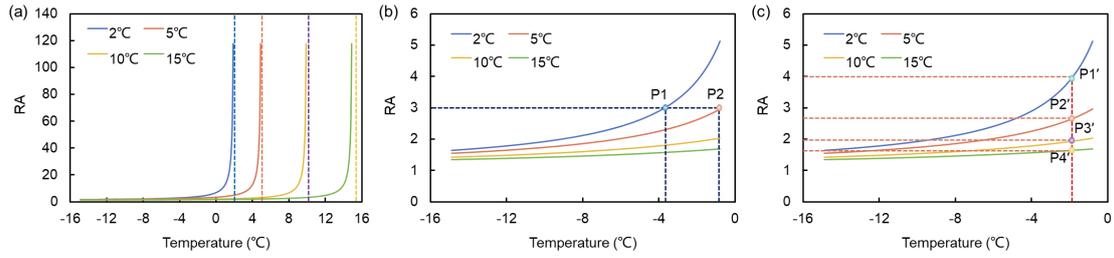

**Fig. 12.** The theoretical results of the variation of RA with substrate surface temperature. (a) RA increases slowly when the temperature is far from $T_{dew}$. As the substrate surface temperature approaches $T_{dew}$, RA increases sharply. Four cases of $T_{dew}$ being 2°C, 5°C, 10°C and 15°C are shown. (b) The enlarged curves of four cases at the range from -16°C to 0 °C. RA increases smoothly in the stage for most cases. For different cases, obtaining the same RA needs lower temperature for the case with low dew temperature. (c) For a fixed substrate surface temperature, RA decreases as the dew temperature increases.

In the temperature range considerably distant from $T_{dew}$, the variation of the ratio RA indicates an approximately linear relationship, as depicted in Fig. 12b. Specifically, within the temperature range of -16°C to 0°C, the curves exhibit a smooth increase. In Fig. 12b, a comparison of condensation temperatures is presented for the same RA value (RA = 3). Two distinct points are labeled: $P_1$ corresponds to $T_{dew}$ = 2 °C, and $P_2$ corresponds to $T_{dew}$ = 5 °C. Notably, the temperature associated with $P_2$ is higher than that of $P_1$. This indicates that for the case of high dew temperature, the substrate temperature needs to be higher than that of low dew temperature when one aims to keep the same dry zone area or dry zone ratio. In Fig. 12c, at a constant substrate surface temperature, the RA value is greater in cases with the lowest dew temperature. As the dew temperature increases, the RA value decreases. We can observe four distinct points denoted as $P_1'$, $P_2'$, $P_3'$, and $P_4'$, each representing the RA values for cases with various dew temperatures at a consistent substrate surface temperature. It follows a clear pattern that $P_1' > P_2' > P_3' > P_4'$, mirroring the order of $T_{dew}$ values: $T_{dew1} < T_{dew2} < T_{dew3} < T_{dew4}$. The comparison of experimental data and the theoretical curve is presented in Fig. 13, which shows data for cases involving 1 μl (a), 2 μl (b), 3 μl (c), and 4 μl (d). In this figure, the dots and dotted line represent the experimental data and theoretical curve, respectively. The results demonstrate a high level of agreement between the theoretical model and the experimental data. This alignment underscores the effectiveness of the theoretical model in describing the observed phenomena.

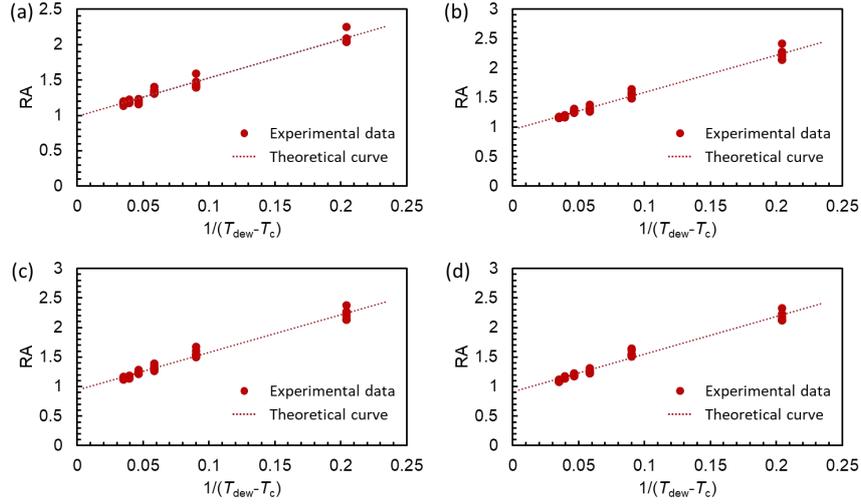

**Fig. 13.** The comparison of experimental data and the theoretical curve. (a)-(d) The cases of 1 μl (a), 2 μl (b), 3 μl (c) and 4 μl (d). The theoretical curve agrees with experimental data well.

## 5. Conclusions

In summary, we conducted an experimental and theoretical investigation into the anti-condensation properties of DG droplets on a cold hydrophobic surface. Our results revealed the formation of a dry zone, within which water vapor condensation is effectively suppressed, around the DG droplet during the cooling of the substrate surface. When the solid surface temperature decreases, the area of dry zone diminishes. This reduction is attributed to the declining solid surface temperature, which causes the boundary of the condensation area to approach the DG droplet, resulting in a contraction of the drying area. To quantify this phenomenon, we introduced the parameter RA, representing the ratio of the dry-zone radius to the radius of the DG droplet. Our findings indicate that RA exhibits a hyperbolic relationship with the decreasing substrate surface temperature. RA remains nearly constant during a brief period of substrate surface cooling, suggesting that cooling time has a limited effect on RA within a specific period. Moreover, at higher substrate surface temperatures, RA decreases with an increase in DG droplet volume, while it remains relatively stable at lower substrate surface temperatures. To provide further insights, we propose a simple yet effective theoretical model that elucidates the relationship between substrate surface temperature and RA. The theoretical predictions closely align with the experimental data, underscoring the effectiveness of our proposed model.

Acknowledgments:

This work was supported by the National Natural Science Foundation of China (No. 12202461).